# Plasmonic Gold Helices for the visible range fabricated by oxygen plasma purification of electron beam induced deposits

*Caspar Haverkamp\* [1, 2], Katja Höflich[1, 2], Sara Jäckle[1, 2], Anna Manzoni[1,] Silke Christiansen[1, 2, 3]*

[1]Helmholtz-Zentrum Berlin für Materialien und Energie GmbH, Hahn-Meitner-Platz 1, 14109 Berlin, Germany

[2]Christiansen Research Group, Max Planck Institute for the Science of Light, Günther-Scharowsky-Str. 1, 91058 Erlangen, Germany

[3]Physics Department, Freie Universität Berlin, Arnimallee 14, 14195 Berlin, Germany

**Abstract**

Electron beam induced deposition (EBID) currently provides the only direct writing technique for truly three-dimensional nanostructures with geometrical features below 50 nm. Unfortunately, the depositions from metal-organic precursors suffer from a substantial carbon content. This hinders many applications, especially in plasmonics where the metallic nature of the geometric surfaces is mandatory. To overcome this problem a post-deposition treatment with oxygen plasma at room temperature was investigated for the purification of gold containing EBID structures. Upon plasma treatment, the structures experience a shrinkage in diameter of about 18 nm but entirely keep their initial shape. The proposed purification step results in a core-shell structure with the core consisting of mainly unaffected EBID material and a gold shell of about 20 nm in thickness. These purified structures are plasmonically active in the visible wavelength range as shown by dark field optical microscopy on helical nanostructures. Most notably, electromagnetic modeling of the corresponding scattering spectra verified that the thickness and quality of the resulting gold shell ensures an optical response equal to that of pure gold nanostructures.

Plasmonic deals with strongly geometry and material dependent collective excitations of the free-electron gas in metallic nanostructures. The resonant behavior of such plasmonic nanostructures allows for extreme light localization and, in case of chiral three-dimensional nanostructures like helices, also for the manipulation of circular polarization states[1] and possibly even non-linear effects.[2]

These geometrically highly demanding truly three-dimensional nanostructures can be fabricated using electron beam induced deposition (EBID).[3] EBID is a direct writing method, a central advantage compared to indirect methods like etching[4] or lithography.[5] The phenomenon that material can be deposited by the impact of an electron beam has been known since the early days of electron microscopy. It often occurs as unwanted sample contamination during the imaging process.[6] Nowadays, there exists a variety of precursors which can be used to intentionally deposit different kinds of material by electron beam induced dissociation.[3] Hereby, a precursor gas is locally injected through a needle into the vacuum chamber of an electron microscope. The electron beam decomposes the molecules of the precursor into a volatile part which is pumped out of the chamber, and a non-volatile part which forms the deposits onto the substrate.[3] Due to the small spot sizes of the electron beam in combination with its flexible control, the EBID process enables the fabrication of structures with nanometer lateral resolution[7] and of complex three-dimensional shapes.[8] Today, the most common applications of EBID range from photo mask repair[9] to the fabrication of tips with high aspect ratio for atomic force microscopy[10] or the deposition and optimization of conductive structures, e.g. for building nano electrodes.[11] However, even beyond electrical applications, the high lateral resolution in combination with the three-dimensional capability render EBID a promising fabrication tool in nano-optics, and especially plasmonics.

For the deposition of metallic structures, precursors, based on metal-organic compounds, are commonly used, resulting in composite deposits consisting of a carbonaceous matrix in which metal crystals are embedded.[12] The high carbon content leads to unwanted high resistivities,[12,13] as well as to a drastically changed optical behavior,[14] of the EBID material compared to the pure metal. One exception is the purely inorganic precursor $PF_3AuCl$ which enables carbon-free deposits with resistivities of only ten times the value of bulk gold.[15] However, the containing fluorine and chlorine are strong etching agents having a negative impact on the vacuum chamber itself[16] and the resulting deposits consists of small particles,[16] making the fabrication of three-dimensional structures difficult.

Another way to reduce the amount of carbon and further impurities to a certain extent is provided by tuning the deposition parameters using e.g. higher electron beam currents.[17] Yet, the thereby achievable improvement is limited and the high beam currents are undesired for the achievement of a minimal resolution.[18]

Since the strong resonances in plasmonics rely on excitations of the free electron, only present in pure metals, the unavoidable carbon content is the major drawback of fabrication using EBID. Hence, there is a variety of purification techniques to reduce the carbon content. Up to now most techniques rely on in-situ substrate heating during the deposition process[19] or post-treatments of the structures by annealing in different oxidizing atmospheres[20,21], electron irradiation or laser treatment of the structures.[22] While these studies show good results for the purification of mainly planar deposits, the purification of complex three-dimensional EBID nanostructures still remains a significant challenge.

First attempts towards the purification of complex gold-based EBID structures while preserving their shape have been made. By using only moderate temperatures around 175°C and ozone as an oxidizing agent a complete purification could be achieved, even though the shape could not be perfectly preserved.[8] Further investigations on vertical standing nanopillars using the precursor $Me_2$-Au-tfa applied an *in-situ* electron beam treatment in combination with a postdeposition treatment and found areas of pure gold crystals inside the structures[23], similar to earlier results in an environmental scanning electron microscope.[24] However, in both cases no closed metal surface was present onto the nanostructures.

Such a plasmonically active surfaces can in principal be achieved using the EBID structures as a template which is coated by metal through thermal evaporation.[25] Where hereby a conformal coating of complex geometries is difficult to achieve due to shadowing effects during the coating.

So far, concerning plasmonic applications, nanostructures written using an ion beam and platinum containing precursors without any additional treatment showed a promising optical performance e.g. arrays of helices showed a circular dichroism in the visible range.[26] However, the optical description of composite nanostructures is challenging.[14]

Furthermore gold, instead of platinum, would represent an ideal choice for plasmonics nanostructures since they combine strongly resonant features in the visible range with full chemical stability.

To directly achieve plasmonically active metal surfaces in combination with shape preservation, a simple and fast oxygen plasma cleaning step at room temperature is investigated. First, the quality of purification is addressed by measuring the resistivity of planar structures and the atomic composition with EDX. The geometrical modification of the structure by the oxygen plasma is examined by REM and TEM measurements on single pillars. Next, the purification of more complex three dimensional structures, in this study gold helices with three pitches, is investigated. Finally, the plasmonic activity of the purified helices is examined using dark field scattering spectra and compared to finite element modeling of geometrically equivalent pure gold helices. One important aspect in terms of purification for optical applications is the strongly restricted penetration of electro-magnetic fields at optical frequencies, called the skin-effect. Typical skin depths of gold in the visible range are of the order of 15 nm[27] what directly implies that not more than the realization of a gold shell with this critical thickness is necessary.

**Results and Discussion**

Different geometries are fabricated by EBID using the metal-organic precursor $Me_2Au(acac)$ and post treated by cold oxygen plasma in a table-top laboratory plasma cleaning system: (I) Single standing pillars to study the shrinking and purification depth of the process, (II) planar structures *i.e.* lines and rectangles to investigate the electrical properties of the purified material as well as the atomic composition, and (III) helices with three pitches to prove the shape preservation even for complex structures and the plasmonic activity.

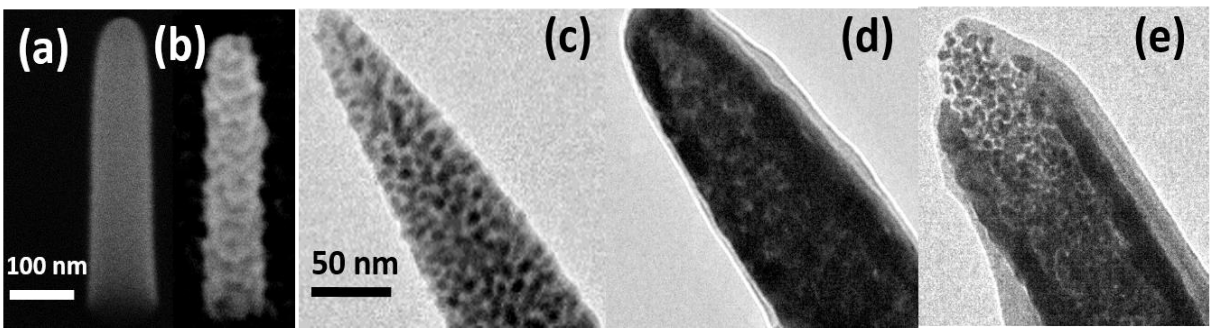

**Figure 1**. Scanning electron micrographs in tilted view (38°) of single EBID pillar deposited using the precursor Me2Au(acac) on silicon substrate (a) before and (b) after a plasma treatment of 2 min which results in a reduction in diameter and height of 18 nm. Transmission electron micrographs of a pillar (c) as deposited, (d) purified and, (e) after an additional cross-section cut to increase the visibility of the core-shell structure. The 100 nm scale bar is valid for (a) and (b) and the 50 nm scale bar for (c), (d), and (e). The bright layers onto the visible pure gold layer are caused by carbon deposition from residual gases present in the vacuum chamber of the TEM.

First the influence of the plasma treatment on the geometry of single pillars is investigated. The scanning electron micrographs in Figure 1 (a) – (b) show the pillar geometry before and after treatment with oxygen plasma for two minutes. The pillar experiences a reduction in volume when exposed to the oxygen plasma, accompanied by a significant increase of surface roughness. A systematic investigation of EBID pillars of varying diameter but equal heights shows, that the reduction in diameter by the plasma treatment is independent from the initial diameter (see Figure 2). Instead, all pillars exhibit the same overall reduction of about 18 nm, suggesting that the purification process has a certain constant penetration depth. A possible explanation is that the plasma has a limited penetration depth, forming a closed gold shell beyond which no further purification and, therefore, no further shrinking occurs. The reduction in size is caused by the oxidation of carbon atoms in the EBID deposit by the impact of oxygen radicals forming volatile species like e.g. CO, $CO_2$.[28]

However, the stability of the shape showed a significant dependence on the diameter. While pillars with diameters larger than 60 nm reliably maintain their shapes, decreasing diameters lead to increasing deformation probability. This observation suggests a critical thickness of the EBID pillars, below which they are not stabile against the heat induced during the plasma treatment (see supporting information).

To gain detailed insights into the mechanism of oxygen plasma purification of the composite EBID material and to verify the proposed core shell structure, transmission electron microscopy (TEM) of the pillars as deposited and after plasma treatment is performed. The micrographs in Figure 1 (c) – (e) show the morphology change occurring during the purification process. The pillar in the as-deposited state Fig 1(c) has the typical composite structure of gold crystals (dark spots) equally distributed within the bright carbonaceous matrix. In the purified pillar Figure 1(d) the core-shell structure can hardly been distinguished due to the decreased transparency for the electron beam upon purification. Therefore, thinning with a focused ion beam milling was employed to unveil the core-shell structure more clearly Figure 1(e). This cross-sectional view reveals the closed outer shell of high gold content with a thickness of about 20 nm, and the inner core of unpurified EBID material. The bright layers onto the visible pure gold layer are caused by carbon deposition from residual gases present in the vacuum chamber.

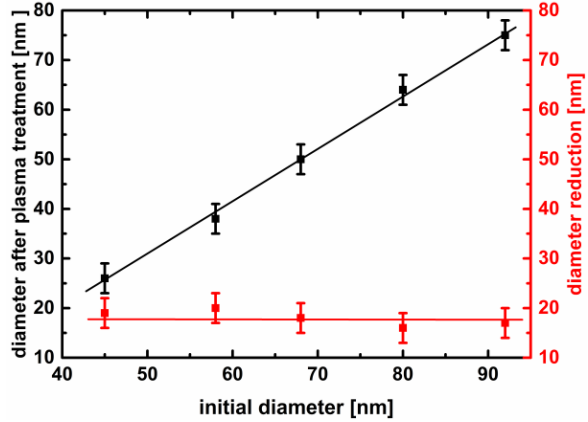

**Figure 2.** Diameter reduction of EBID pillars for different starting thicknesses after purification of 2 min in oxygen plasma. Lines are only for guidance.

A more detailed investigation of the time evolution of purification (see supporting information) propose the top down nature of the purification, or, in the case of pillars, the out-to-inside. This is in agreement with earlier studies on EBID purification of platinum precursor in oxygen atmosphere under electron radiation,[28] where the top-down mechanism is attributed to a low permeability coefficient and a high chemisorption of oxygen, which results in a purification restricted close to the surface. In contrast to the treatment of platinum deposition in water vapor with a higher permeability coefficient and a lower chemisorption a bottom-up purification has been noted.[29] The observed out- to inside purification in this study is a first hint that the permeability coefficient of the radicals is comparably small in the case of oxygen plasma applied for gold-containing deposits. The growth of gold layers is accompanied by Ostwald ripening, resulting in the formation of islands.[30] Depending on the temperature a closed layer of gold occurs between 10 – 20 nm,[31] suggesting that the purification with oxygen plasma is self-limiting up to the point where a closed gold layer has formed, hindering further penetration of the oxidizing species.

As mentioned before, the resistivity of EBID material is related to its metal content.[12] To confirm the purification effect of the plasma treatment and to quantify the improvement, the electrical resistivity of the EBID material before and after the treatment is compared. A two point resistance measurement on four EBID bridges between differently spaced gold pads is performed to eliminate the contribution of contact resistance. Figure 3(a) shows an atomic force micrograph along with the extracted height profile of an EBID bridge geometry used for determining the cross-sectional area. Figure 3(b) presents the corresponding dependence of the measured resistance before and after the plasma purification on length/area of the bridge. From the slope the specific resistivity was determined to be 360 (±14) µΩm for the pure EBID material, strongly decreasing to 10 (±2) µΩm for the purified material. In the same manner the contact resistance, extracted from the y-axis intercept, drops from 173 (±45) Ω to 60 (±6) Ω. This large improvement of conductivity is remarkable remembering that only the outer shell of the material is purified while the core still consist of unpurified EBID material. Taking into account the core-shell geometry the deduced specific resistivity only represents an upper limit for the purified EBID material in the shell. If assuming a constant 20 nm thick shell throughout the whole bridge, based on the TEM results, surrounding otherwise unchanged EBID material with the specific resistivity measured before purification, this results in a specific resistivity of 1.1 (± 0.5) µΩm for the shell material. While the specific resistivity of bulk gold (0.02 µΩm[32]) is almost two orders of magnitude smaller compared to the purified shell, thin layers of pure gold typically show lower resistivities compared to the bulk value, e.g. about 0.4 Ωµm for a layer thickness of 37 nm.[33] This behavior can be explained by the strong dependence of resistivity on the number of grain boundaries inside the material.[33] Thin metallic layers typically show reduced grain sizes and thereby a significantly higher density of grain boundaries. This holds true for the plasma purification process, where the layer is assembled from the nm-sized gold grains present in the EBID material (see TEM picture Figure 1(c)). Additionally the strong surface roughness of the purified EBID surface (see Figure 1(b)) actually makes the assumption of a smooth 20 nm layer also an upper estimation to the specific resistivity of the shell material. Certainly also possible is an imperfect purification of the EBID shell still containing traces of carbon and oxygen. As mentioned before, the deposition parameters have an influence on the resulting structure and atomic composition and are not identical for planar and three dimensional deposits. In addition, when depositing planar structures compared to nanostructures, the larger area results in less substrate heating during the process[3] and thus results in a lower metal content.[34] Furthermore the supply of molecules and thus the resulting composition is different.[3] Thus, it can't be guaranteed that the conductivity determined from the planar structures is exactly the same as for the pillars. But since the general structure and atomic constituents of the material are the same, the purification effect for planar structures and pillars should be comparable. And since the measured resistivity is mostly determined by the purified shell, the value should be a good approximation for vertical structures also.

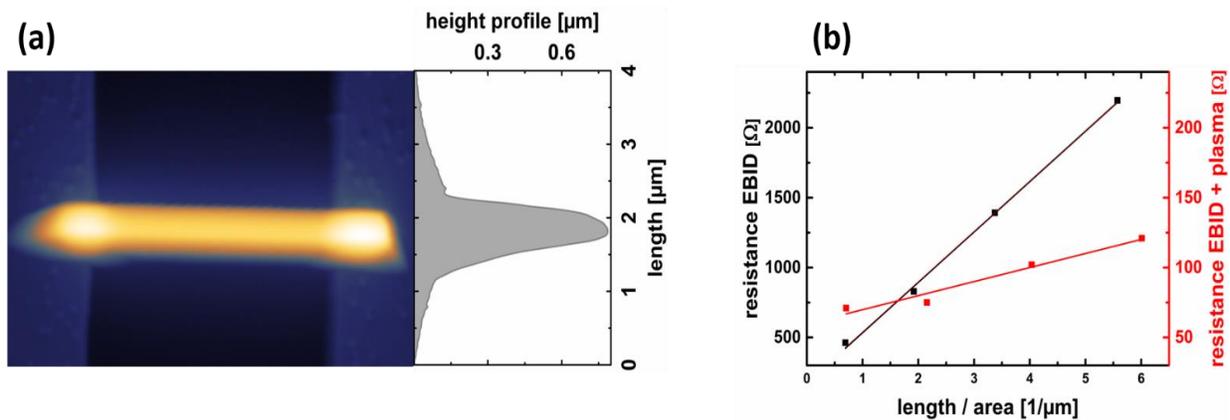

**Figure 3.** (a) Atomic force micrograph and extracted height profile of EBID bridge between two gold pads. (b) Resistance over length/area of EBID bridges measured between four different distances before (black) and after (red) plasma purification.

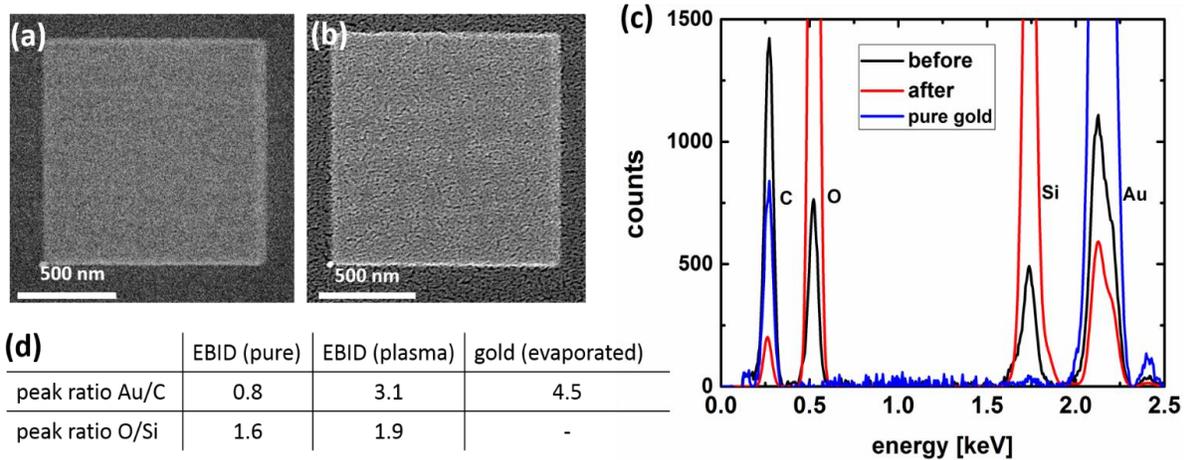

**Figure 4**. Scanning electron micrographs of the EBID pad (a) as deposited and (b) plasma purified, (c) EDX spectra of EBID pad as deposited (black curve), purified (red curve) and pure gold (blue curve) for comparison. (d) Peak ratios of gold to carbon for all three cases and oxygen to silicon for the EBID material before and after purification.

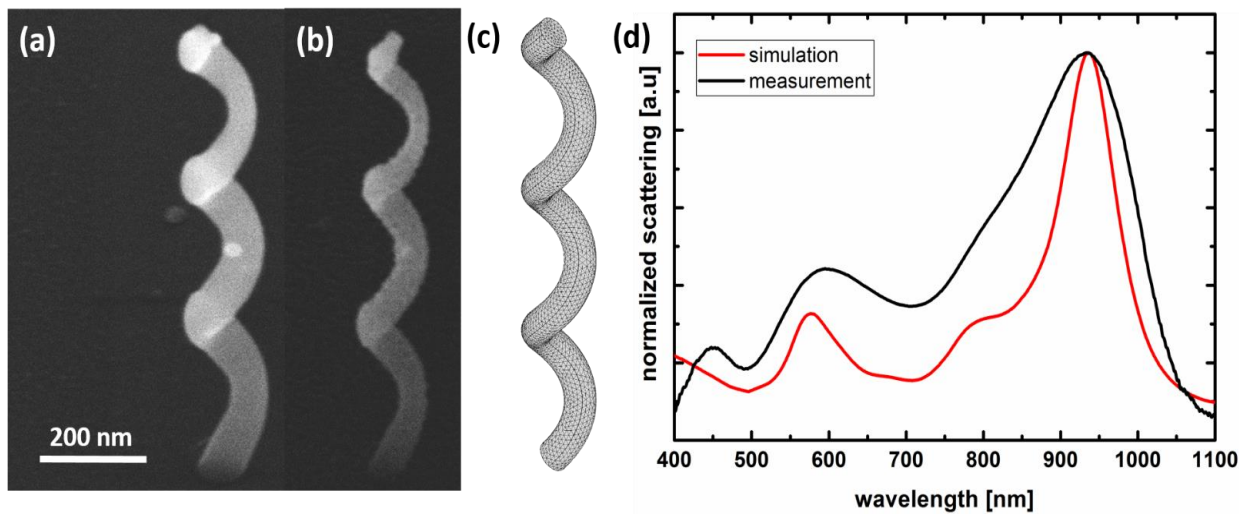

**Figure 5.** Scanning electron micrographs in tilted view of helix with three pitches before (a) and (b) after plasma purification. (c) Mesh plot of the corresponding modelled helix, (d) comparison of the measured and simulated scattering spectra.

The presented results, volume shrinking, the formation of a core-shell structure as well as the improvement of electrical conductivity propose that carbon is removed from the EBID composite. While these methods provide for an indirect access to the amount of gold and carbon, energy dispersive X-ray diffraction (EDX) provides a direct measurement of the atomic composition. However, the determination of the exact amount of light elements using EDX is rather difficult. This is true especially for the case of carbon constituting a typical contamination of residual gases inside the vacuum chamber. Therefore, the quantification presented here relies on peaks ratios of Au:C after subtracting the background instead of absolute amounts.[29]

Figures 4 (a) and (b) show scanning electron micrographs of the investigated planar deposits before and after oxygen plasma treatment. As in the case of pillars a distinct surface roughening is observed; even cracks and holes are visible. The height of the purified pad was 12 nm in average. Thus, assuming that the thickness of the purified layer of the planar structure is comparable with the one of the pillar the plasma cleaned pad should consist of purified material only. A pure gold sample with 250 nm of gold evaporated onto a silicon wafer was used for comparison. The corresponding EDX spectra and peak ratios are presented in Figure 4 (c) and (d). The fact that the EDX spectrum of the pure gold sample (blue curve) shows no signal of the underlying silicon substrate proves that the signal originates from gold material only. However, as expected even the pure gold sample exhibits a distinct carbon signal resulting from the residual gases inside the chamber. Comparing the peak ratios of gold to carbon in the pure EBID material with the purified material, an increase from 0.8 to 3.1 is observed. The silicon peak of the substrate is increased in the EDX spectra of the plasma treated sample because of the thinner layer of material on the one hand and because of the formation of voids on the other hand. The increase of the oxygen peak in Figure 4 (d) can be mainly attributed to the larger influence of the substrate. The EDX measurements show that the reduction in diameter and the improvement of the electrical conductivity can be attributed to a substantial effective removal of carbon atoms from the EBID material.

To demonstrate that the plasma cleaning step indeed purifies complex structures to such an extent that they can optically compete with pure gold, helices with three pitches were fabricated and plasma treated. Figure 5 (a)-(c) shows scanning electron micrographs of a helix as-deposited and after purification, along with a mesh plot of the helix modeled. The shape of the helix is well maintained by the purification step and no distortions occurred. The dimensions for the modeled helix are carefully extracted from the SEM images, under consideration of the tilted view of 38°. To study the optical response of the purified helix, dark field scattering spectra are recorded and compared to scattering efficiencies from electrodynamic modeling of a pure gold helix (see Figure 5(c)).

Besides the expected significantly larger broadening of the resonances due to the unavoidable higher losses in the experiment, possible caused by grain boundaries, surface roughness and traces of carbon, the overall agreement is remarkable. Both spectra show two distinct strong resonances and, interestingly, one small resonance in-between, visible as a shoulder, because of less broadening more pronounced in case of the experimentally determined spectrum. The experimentally determined position of the short wavelength resonance around 570 nm is slightly red-shifted compared to the simulated results. The long wavelength resonance at 950 nm instead matches almost perfectly with the simulation and in the same manner the shoulder around 790 nm is well-reproduced in the simulation. The only deviation from the modeling is one additional peak around 450 nm showing up in the experimental spectrum only. This could possibly originate from purified deposits around the helix or surface roughness induced by the purification. The overall agreement between experiment and theory proves that an optical description in terms of pure gold nanostructured is justified for the presented purification routine. This shows, for the first time, pasmonic gold helices having resonance in the visible range. To further study the observed resonant features a systematic investigation using adapted polarization states like e.g. circular polarization in combination with corresponding simulations is under current investigation.

**Conclusions**

In this work a fast and simple oxygen plasma treatment for nanostructures fabricated with electron beam induced deposition has been presented. In contrast to former purification studies, which focused mainly on planar deposits, the presented purifications method was confirmed to be successful in case of complex three dimensional helical nanostructures. While the diameters were reduced, the shapes of the helices were entirely maintained. TEM investigation of the cross section of EBID pillars revealed that the purification results in a core-shell structure whereby the core consists of mostly unaffected EBID composite with a 20 nm thick purified shell. The quality improvement of the material was investigated by electrical and EDX measurements on planar deposits showing that the shell exhibits a drastically high content of gold, compared to the pure EBID material. The optical investigation employing dark field reflection measurements displayed two distinct resonances of a purified helix lying in the visible and near infrared region. These resonances show an excellent agreement with simulated scattering spectra of geometrically equivalent gold helices from finite element modeling. This result proves that the arising gold surface is both, plasmonically active and has a sufficient thickness to neglect the remaining composite core in the optical response. In summary, the subsequent plasma cleaning step provides a simple, fast and reliable tool for the direct surface purification of complex-shaped EBID nanostructures and thereby the missing link for an easy utilization of EBID in plasmonics. Hence, the post deposition plasma treatment in combination with the 3D flexibility of EBID paves the way for an efficient fabrication of three-dimensional gold nanostructures of unprecedented flexibility and resolution for various plasmonic applications.

**Methods**

All EBID depositions were carried out in an FEI Strata dual beam system using the gold containing metal-organic precursor $Me_2Au(acac)$, which is inserted inside the chamber through an integrated gas injection system (GIS). The temperature of the precursor reservoir was fixed at 34

°C throughout all growing processes. The electron beam movement was controlled by directly addressing a patterning board using stream files containing coordinates and dwell times for each pixel.

Whereas the deposition parameters electron energy (15 kV), beam current (203 pA) and working distance (5 mm) were kept constant for all geometries. To achieve three dimensional constructs instead of planar deposits long dwell times and/or small pixel spacings are necessary. Thus, for the planar structures, a dwell time of 1 µs and a pixel spacing of 5 nm were used while for the pillars and helices the pixel spacing had to be reduced to 1 nm and the dwell time was increased to 1.2 ms. To realize helices with similar pitch heights for each turn the dwell time has been increased to 1.4 ms and 1.8 ms for the second and third turn, respectively.[26] To avoid any lamella preparation and transfer process for transmission electron microscopy (TEM), pillars were directly grown on an Omniprobe TEM grids. Only a thinning by a focused gallium beam at 30 kV energy with a beam current of 10 pA was applied to picture the cross section the purified pillars. All TEM images were acquired with a Gatan Orius CCD-Camera inside a CM12 (Phillips) at an accelerating voltage of 120 KV. Except for the TEM grids, all other depositions were done on a silicon substrate with a 300 nm thermally grown oxide. For electrical characterization pure gold pads of 100 nm height with a linearly increasing distance from 0.7 to 3.7 µm onto a 10 nm adhesion layer from titanium were evaporated through a lithographic mask, using an electron beam evaporator type Balzer PLS 570 with base pressure around $10^{-6}$ mbar. Two point resistance measurements were performed using a Keithley SCS 4200 semiconductor characterization system using currents < 1 mA to prevent Joule heating during the measurement. From the linear I-V curves the resistance was extracted (see in Supporting Information). The topographic profiles of the bridges were measured before and after purification using an atomic force microscope of type SmartSPM™ 1000 from AIST-NT. The material composition was determined by energy-dispersive X-ray (EDX) measurements onto square EBID pads with a side length of 1 µm in a Tescan Lyra 3 dual beam microscope equipped with a Bruker EDX Quantax system. All spectra were taken in spot mode with an acquisition time of 60 seconds at an accelerating voltage of 5 kV. As reference samples a gold layer of 250 nm has been evaporated on silicon substrate.

Finally, the purification itself was performed using cold oxygen plasma in a table-top laboratory plasma cleaning system (zepto – diener electronic) operating at 40 kHz. Pressure and input power were optimized to 0.3 mbar and 70 W, respectively, providing an optimal compromise between minimizing structural damage under maximal purification. Scattering spectra were recorded using a Zeiss Axio Imager optical microscope under dark field configuration. The sample was illuminated with unpolarized light of a halogen lamp through a 100x objective (NA 0.75). The scattered light is collected with the same objective and out-coupled through a 400 µm optical fiber to a Horiba iHR 320 spectrometer. Hereby, the experimentally detected signal is proportional to the scattering efficiency of the helix and the light intensity of the source.[35]

Therefore all spectra were normalized by subtracting the background and dividing by the source spectra.

$$I = \frac{I_{scattering} - I_{dark}}{I_{source}}$$

To compare the experimentally retrieved spectra finite element modeling (FEM) was applied, using a three-dimensional spherical simulation domain with perfectly matched layers. The simulation was set up according to earlier models[36] and verified using Mie calculations as the analytic solution for a single sphere. The wavelength was parametrized using Matlab scripting which extracted absorption efficiencies as well directionally dependent scattering efficiencies (forward, backward, side). The geometrical parameters of the gold helix were retrieved from SEM images after purification. Given the fact that the gold shell thickness after purification being above the skin depth of gold at optical frequencies the permittivity of the helices was assumed to be the one of pure gold.[37]

**Corresponding Author**

* E-Mail: caspar.haverkamp@helmholtz-berlin.de

**Author Contributions**

All authors have given approval to the final version of the manuscript.

**Acknowledgements**

The authors acknowledge funding from the Helmholtz Association within the Helmholtz Postdoc Program, financial support by the European Union Seventh Framework Programme (FP7/2007-2013) under the Grant Agreement no 280566, project UnivSEM and from the EU COST Action CM1301 'CELINA'.

*References*

(1) Gansel, J. K.; Thiel, M.; Rill, M. S.; Decker, M.; Bade, K.; Saile, V.; von Freymann, G.; Linden, S.; Wegener, M. Gold Helix Photonic Metamaterial as Broadband Circular Polarizer. *Science* **2009**, *325*, 1513–1515.

(2) Rodrigues, S. P.; Lan, S.; Kang, L.; Cui, Y.; Cai, W. Nonlinear Imaging and Spectroscopy of Chiral Metamaterials. *Adv. Mater.* **2014**, *26*, 6157–6162.

(3) Utke, I.; Hoffmann, P.; Melngailis, J. Gas-Assisted Focused Electron Beam and Ion Beam Processing and Fabrication. *J. Vac. Sci. Technol. B* **2008**, *26*, 1197–1276.

(4) Ranade, R. M. Reactive Ion Etching of Thin Gold Films. *J. Electrochem. Soc.* **1993**, *140*, 3676–3678.

(5) Wu, B.; Kumar, A. Extreme Ultraviolet Lithography: A Review. *J. Vac. Sci. Technol. B* **2007**, *25*, 1743–1761.

(6) Hren, J. J. Barriers to AEM: Contamination and Etching. In *Principles of Analytical Electron Microscopy*; Joy, D.C.; Alton, D.; Romig Jr., J.; Goldstein, I., Ed.; Springer Science+Business Media: New York, 1986; pp. 353–374.

(7) Kouwen, L. Van; Botman, A.; Hagen, C. W. Focused Electron-Beam-Induced Deposition of 3 nm Dots in a Scanning Electron Microscope. *Nano Lett.* **2009**, *9*, 3–6.

(8) Höflich, K.; Yang, R.; Berger, A. The Direct Writing of Plasmonic Gold Nanostructures

by Electron Beam Induced Deposition. *Adv. Mater.* **2011**, *23*, 2657–2661.

(9) Edinger, K.; Becht, H.; Bihr, J.; Boegli, V.; Budach, M.; Hofmann, T.; Koops, W.P.; Kuschnerus, P.; Oster, J.; Spies, P.; Weyrauch, B. Electron-Beam-Based Photomask Repair. *J. Vac. Sci. Technol. B Microelectron. Nanom. Struct.* **2004**, *22*, 2902–2906.

(10) Wendel, M.; Lorenz, H.; Kotthaus, J. P. Sharpened Electron Beam Deposited Tips for High Resolution Atomic Force Microscope Lithography and Imaging. *Appl. Phys. Lett.* **1995**, *67*, 3732–3734.

(11) Brintlinger, T.; Fuhrer, M. S.; Melngailis, J.; Utke, I.; Bret, T.; Perentes, A.; Hoffmann, P.; Abourida, M.; Doppelt, P. Electrodes for Carbon Nanotube Devices by Focused Electron Beam Induced Deposition of Gold. *J. Vac. Sci. Technol. B* **2005**, *23*, 3174–3177.

(12) Van Dorp, W. F.; Hagen, C. W. A Critical Literature Review of Focused Electron Beam Induced Deposition. *J. Appl. Phys.* **2008**, *104*, 081301.

(13) Koops, H. W. P. High-Resolution Electron-Beam Induced Deposition. *J. Vac. Sci. Technol. B Microelectron. Nanom. Struct.* **1988**, *6*, 477–481.

(14) Woźniak, P.; Höflich, K.; Brönstrup, G.; Banzer, P.; Christiansen, S.; Leuchs, G. Unveiling the Optical Properties of a Metamaterial Synthesized by Electron-Beam-Induced Deposition. *Nanotechnology* **2015**, *27*, 025705.

(15) Utke, I.; Hoffmann, P.; Dwir, B.; Leifer, K.; Kapon, E.; Doppelt, P. Focused Electron Beam Induced Deposition of Gold. *J. Vac. Sci. Technol. B Microelectron. Nanom. Struct.* **2000**, *18*, 3168–3171.

(16) Botman, A.; Mulders, J. J. L.; Weemaes, R.; Mentink, S. Purification of Platinum and Gold Structures after Electron-Beam-Induced Deposition. *Nanotechnology* **2006**, *17*, 3779–3785.

(17) Weber, M. Electron-Beam Induced Deposition for Fabrication of Vacuum Field Emitter Devices. *J. Vac. Sci. Technol. B* **1995**, *13*, 461–464.

(18) Beaulieu, D.; Ding, Y.; Wang, Z. L.; Lackey, W. J. Influence of Process Variables on Electron Beam Chemical Vapor Deposition of Platinum. *J. Vac. Sci. Technol. B Microelectron. Nanom. Struct.* **2005**, *23*, 2151–2159.

(19) Blauner, P. G. Focused Ion Beam Induced Deposition of Low-Resistivity Gold Films. *J. Vac. Sci. Technol. B Microelectron. Nanom. Struct.* **1989**, *7*, 1816–1818.

(20) Botman, A.; Mulders, J. J. L.; Hagen, C. W. Creating Pure Nanostructures from Electron-Beam-Induced Deposition Using Purification Techniques: A Technology Perspective. *Nanotechnology* **2009**, *20*, 372001.

(21) Elbadawi, C.; Toth, M.; Lobo, C. J. Pure Platinum Nanostructures Grown by Electron Beam Induced Deposition. *ACS Appl. Mater. Interfaces* **2013**, 1–5.

(22) Stanford, M.G.; Lewis, B. B. .; Noh, J.H.; Fowlkes, J.D.; Roberts, N.A.; Plank, H.; Rack, P. D. Purification of Nanoscale Electron-Beam-Induced Platinum Deposits via a Pulsed Laser-Induced Oxidation Reaction. *ACS Appl. Mater. interfaces* **2014**, 21256–21263.

(23) Belić, D.; Shawrav, M. M.; Gavagnin, M.; Stöger-Pollach, M.; Wanzenboeck, H. D.; Bertagnolli, E. Direct-Write Deposition and Focused-Electron-Beam-Induced Purification of Gold Nanostructures. *ACS Appl. Mater. Interfaces* **2015**, *7*, 2467–2479.

(24) Mølhave, K.; Madsen, D. Solid Gold Nanostructures Fabricated by Electron Beam Deposition. *Nano Lett.* **2003**, *2*, 1–5.

(25) Höflich, K.; Becker, M.; Leuchs, G.; Christiansen, S. Plasmonic Dimer Antennas for Surface Enhanced Raman Scattering. *Nanotechnology* **2012**, *23*, 185303.

(26) Esposito, M.; Tasco, V.; Cuscuna, M.; Todisco, F.; Benedetti, A.; Tarantini, I.; Giorgi, M. De; Sanvitto, D.; Passaseo, A. Nanoscale 3D Chiral Plasmonic Helices with Circular Dichroism at Visible Frequencies. *ACS Photonics* **2015**, *2*, 105–114.

(27) Becker, J. Light-Scattering and -Absorption of Nanoparticles. In *Plasmons as sensors*; Springer-Verlag Berlin: Heidelberg, 2012; pp. 5–38.

(28) J.D. Fowlkes; B. Geier; B.b. Lewis; P.D. Rack; M.G. Standfird; R.Winkler; H.Plank. Electron Nanoprobe Induced Oxidation : A Simulation of Direct-Write Purification. *Phys. Chem. Chem. Phys.* **2015**, *17*, 18294–18304.

(29) Geier, B.; Gspan, C.; Winkler, R.; Schmied, R.; Fowlkes, J. D.; Fitzek, H.; Rauch, S.; Rattenberger, J.; Rack, P. D.; Plank, H. Rapid and Highly Compact Puri Fi Cation for Focused Electron Beam Induced Deposits: A Low Temperature Approach Using Electron Stimulated $H_2O$ Reactions. *J. Phys. Chem. C* **2014**.

(30) Listvan, M. A. Direct Observations of Small Gold Clusters and in Situ Cluster Growth by Stem. *J. Mol. Catal.* **1983**, *20*, 265–278.

(31) Schaub, A.; Slepička, P.; Kašpárková, I.; Malinský, P.; Macková, A.; Svorčík, V. Gold Nanolayer and Nanocluster Coatings Induced by Heat Treatment and Evaporation Technique. *Nanoscale Res. Lett.* **2013**, *8*, 249.

(32) Munoz, R. C.; Vidal, G.; Mulsow, M.; Lisoni, J. G.; Arenas, C.; Concha, A.; Mora, F.; Espejo, R.; Esparza, R.; Haberle, P. Surface Roughness and Surface-Induced Resistivity of Gold Films on Mica: Application of Quantitative Scanning Tunneling Microscopy. *Phys. Rev. B* **2000**, *62*, 4686–4697.

(33) Zhang, Q. G.; Cao, B. Y.; Zhang, X.; Fujii, M.; Takahashi, K. Influence of Grain Boundary Scattering on the Electrical and Thermal Conductivities of Polycrystalline Gold Nanofilms. *Phys. Rev. B - Condens. Matter Mater. Phys.* **2006**, *74*, 1–5.

(34) Mulders, J.J.L.; Belova, L. M.; Riazanova, A. Electron Beam Induced Deposition at Elevated Temperatures: Compositional Changes and Purity Improvement. *Nanotechnology* **2011**, *22*, 055302.

(35) Brönstrup, G.; Leiterer, C. Jahr, N.; Gutsche, C.; Lysov, A.; Regolin, I.; Prost, W.; Tegude, F. J.; Fritzsche, W.; Christiansen, S. A Precise Optical Determination of Nanoscale Diameters of Semiconductor Nanowires. *Nanotechnology* **2011**, *22*, 385201.

(36) Höflich, K.; Gösele, U.; Christiansen, S. Near-Field Investigations of Nanoshell Cylinder Dimers. *J. Chem. Phys.* **2009**, *131*, 164704.

(37) Johnson, P.b. Christy, R. W. Optical Constants of the Noble Metaks. *Phys. Rev. B* **1972**, *6*, 4370–4379.

**Supporting Information**

I-V Curve of EBID bridge, TEM crossection for different time steps, SEM deformed pillar

**Variation of Plasma Purification Time**

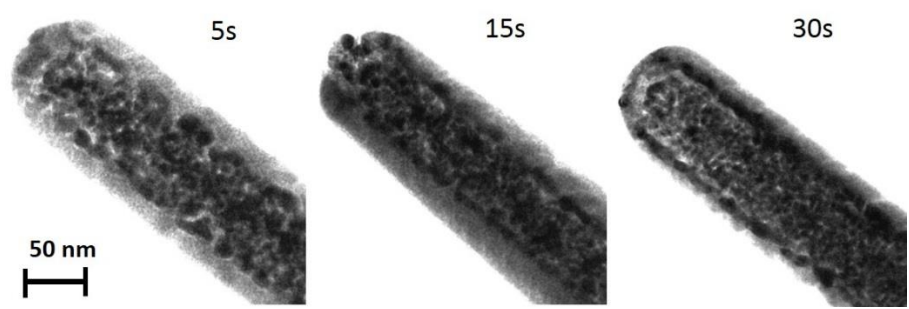

**Figure 1**. Cross section of purified pillars for different time steps.

Figure 1 shows the cross section of EBID pillars for three different time steps. Visible is the formation of the shell at the edge of the pillar while the inner part is unaffected. The bright layers onto the visible pure gold layer are caused by carbon deposition from residual gases present in the vacuum chamber of the TEM.

**Deformation of Thin Pillar**

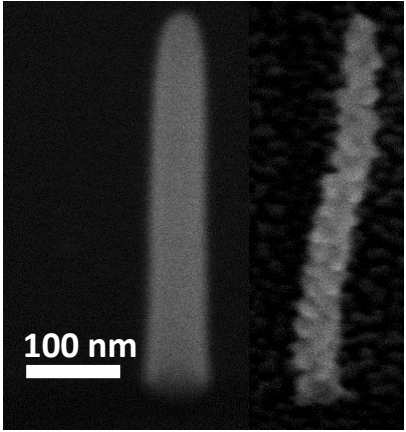

**Figure 2.** Deformation of a thin pillar during

Figure 2 shows the deformation of a thin pillar during the purification in oxygen plasma.

**Measurement of Resistance**

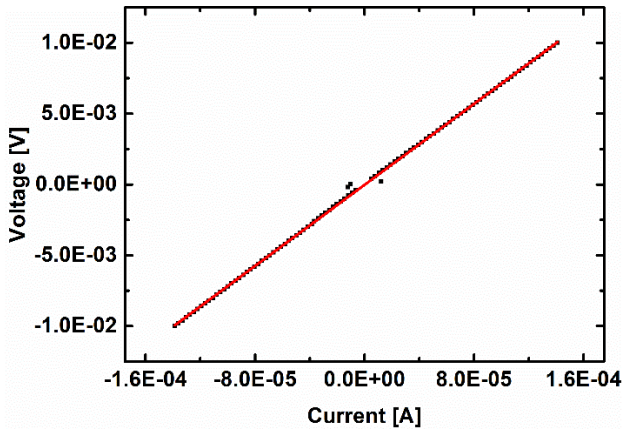

**Figure 3.** Current-Voltage plot of the bridge with a length of 700 nm.

Figure 3 shows an exemplary I-V curve of the first EBID bridge. Due to the linear relation between length and resistance it is possible to extract the specific resistivity $\rho$ as well as the contact resistance between the EBID material and the gold pads using formula

$$R = \rho * \frac{l}{A} + 2 * R_{contact} \quad (1)$$

.The contact resistance in formula 1 takes into account the junction between EBID material and gold pad. The error of the resistivity is the standard deviation of the linear fit.